\documentclass[aps,prd,reprint,superscriptaddress,floatfix]{revtex4-2}

\usepackage{amsmath}
\usepackage{amssymb}
\usepackage{amsthm}
\usepackage{bm}
\usepackage{graphicx}
\usepackage{textcomp}
\usepackage{xcolor}
\usepackage{tikz}
\usetikzlibrary{calc, arrows.meta}
\usepackage{caption}
\usepackage{subcaption}
\usepackage{hyperref}

\newtheorem{theorem}{Theorem}

\tikzset{boximg/.style={remember picture,red,thick,draw,inner sep=0pt,outer sep=0pt}}

\begin{document}

\title{An end-to-end function for the photograph of a Schwarzschild black hole}

\author{Ping Li}
\email[]{Lip57120@huas.edu.cn}
\affiliation{College of Mathematics and Physics, Hunan University of Arts and Sciences, 3150 Dongting Dadao, Changde City, Hunan Province 415000, China}
\affiliation{Hunan Province Key Laboratory Integration and Optical Manufacturing Technology, 3150 Dongting Dadao, Changde City, Hunan Province 415000, China}
\author{Jun Cheng}
\email[]{chengjun@huas.edu.cn}
\affiliation{College of Mathematics and Physics, Hunan University of Arts and Sciences, 3150 Dongting Dadao, Changde City, Hunan Province 415000, China}
\affiliation{Hunan Province Key Laboratory Integration and Optical Manufacturing Technology, 3150 Dongting Dadao, Changde City, Hunan Province 415000, China}

\author{Jiang-he Yang}
\email[]{yjianghe@163.com}
\affiliation{College of Mathematics and Physics, Hunan University of Arts and Sciences, 3150 Dongting Dadao, Changde City, Hunan Province 415000, China}
\affiliation{Center for Astrophysics, Guangzhou University, 230 West Ring Road, Guangzhou, Guangdong Province 510006, China}

\date{\today}

% ---------------------------------------------------------------------------
\begin{abstract}
We show that the photograph of a Schwarzschild black hole can be written in
closed form. The image plane is embedded as a plane tangent to a sphere centred
on the black hole, and the map between a point source and its image point,
which we call the photograph function, is treated as the analytic object.
Explicit expressions are obtained for sources on the celestial sphere at
infinity and for sources in the equatorial plane, without solving for the
intermediate null geodesics. The construction is exact, and it makes the
multivalued branch structure, the relation to the photon sphere and to the
shadow boundary, and the self-similar higher-order images directly readable.
\end{abstract}

\maketitle

% ===========================================================================
\section{Introduction}
\label{sec:introduction}
% ===========================================================================

The Schwarzschild solution was obtained in 1916, soon after the completion of
the general theory of relativity, and was for a long time regarded as a
mathematical curiosity rather than a description of a physical object
\cite{Schwarzschild1916}. Its interpretation changed in the 1930s, when
Oppenheimer and Snyder showed that a collapsing cloud of pressureless dust forms
a singularity hidden behind a surface from which no signal can reach an outside
observer \cite{OppenheimerSnyder1939}, and when the singularity theorems later
placed this result on a general footing \cite{Penrose1965}. Evidence that black
holes exist as astrophysical objects accumulated from the observation of quasars
and active galactic nuclei, from the measurement of stellar orbits around the
Galactic centre \cite{Ghez2008,Genzel2010}, and from the detection of
gravitational waves emitted by merging binary black holes \cite{Abbott2016}. The
most direct visual evidence came from very long baseline interferometry: in 2019
the Event Horizon Telescope (EHT) reported a ring-like image of the supermassive
black hole in M87 \cite{EHT2019L1,EHT2019L6}, and in 2022 a corresponding image
of Sagittarius A* at the Galactic centre \cite{EHT2022L12}. Both images show a
bright ring enclosing a central depression whose size is consistent with the
shadow predicted by general relativity \cite{EHT2021L12,EHT2021L13}.

The theoretical basis of such images was laid well before these observations.
Synge analysed the escape of photons from a gravitationally intense star and
identified the critical impact parameter that separates escaping from captured
rays \cite{Synge1966}. Luminet computed the appearance of a Schwarzschild black
hole illuminated by a background source and showed that the image of the source
is repeated in a sequence of increasingly thin rings \cite{Luminet1979}.
Bardeen described the shadow of the Kerr metric and the spherical photon orbits
that bound it \cite{Bardeen1973}. The common ingredient of these early studies
is the behaviour of rays whose impact parameter is close to a critical value:
such rays wind many times around the black hole before escaping, so that the
whole sky is mapped, with exponentially decreasing scale, onto a narrow
neighbourhood of the shadow edge. This behaviour is usually described in the
strong-deflection limit, in which the deflection angle is expanded
logarithmically about the critical impact parameter. The general form of this
expansion for a spherically symmetric spacetime was obtained by Bozza
\cite{Bozza2002} and by Bozza, Capozziello, Iovane and Scarpetta
\cite{Bozza2001}; it extends earlier analyses of relativistic images by Darwin
\cite{Darwin1959}, Atkinson \cite{Atkinson1965} and Ohanian \cite{Ohanian1987},
and of Einstein rings by Chwolson \cite{Chwolson1924} and Einstein
\cite{Einstein1936}. Applications to Schwarzschild, Reissner--Nordstr\"om and
other compact objects, together with the classification of images and their
magnifications, were given by Virbhadra and Ellis
\cite{VirbhadraEllis2000,VirbhadraEllis2002,Virbhadra2009}, Frittelli, Kling and
Newman \cite{Frittelli2000}, and Eiroa, Romero and Torres \cite{Eiroa2002}.
Reviews of the field may be found in Refs.~\cite{Bozza2010,CunhaHerdeiro2018,
PerlickTsupko2022,Perlick2004}.

In parallel, the exact integrability of null geodesics in black hole spacetimes
has been studied in detail. In the Schwarzschild spacetime the radial equation
becomes a cubic polynomial in the inverse radius, and its inversion is achieved
by elliptic functions. The classical treatment is given in Chandrasekhar's
monograph \cite{Chandrasekhar1983}. Hackmann and L\"ammerzahl provided a
systematic formulation in terms of the Weierstrass $\wp$ function
\cite{Hackmann2008}, and Gibbons and Vyska used the same formalism to describe
Schwarzschild null geodesics completely and to expand the deflection angle to
second order in $M/r_0$ and $M/b$ \cite{GibbonsVyska2012}. In the Kerr
spacetime, the separability of the Hamilton--Jacobi equation established by
Carter \cite{Carter1968}, together with the perturbative framework of Bardeen,
Press and Teukolsky \cite{Bardeen1972} and the formulation of
Cunningham and Bardeen \cite{Cunningham1973}, provides analytic trajectories.
Gralla and Lupsasca have given a complete parametrisation of Kerr null geodesics
in terms of elliptic integrals, expressed through the Mino time
\cite{Mino2003}, which is convenient for practical use \cite{GrallaLupsasca2020a},
and have used it to describe the shape and the observable properties of the
black hole image \cite{GrallaLupsasca2020b,Gralla2021}. Analytic formulae for higher-order images have been derived by
Tsupko \cite{Tsupko2022} and by Bisnovatyi-Kogan and Tsupko
\cite{BisnovatyiKoganTsupko2022,BisnovatyiKoganTsupko2017}, who express the
apparent shape of the higher-order rings of equatorial emitters as explicit
polar curves, and by Aratore, Tsupko and Perlick \cite{Aratore2024}, who obtain
an analytic gap parameter for neighbouring photon rings in a general
spherically symmetric spacetime. A review of critical orbits in a system can be found in the literature \cite{Li2026-1}.

Alongside this analytic work, numerical ray tracing and general relativistic
magnetohydrodynamic simulations have become the standard tool for interpreting
EHT data \cite{EHT2019L1,EHT2022L12}. The photon ring and its substructure, the
most strongly lensed part of the image, have received particular attention.
Johnson and collaborators showed that the sequence of subrings produces a
characteristic signature on long interferometric baselines
\cite{Johnson2020}, and Himwich and collaborators derived the corresponding
polarimetric signature \cite{Himwich2020}. Gralla, Holz and Wald
\cite{GrallaHolzWald2019} developed a systematic description of photon rings,
lensing rings and their observables. Sneppen reformulated the problem in terms
of the linear stability of the photon orbit and showed that successive images
differ by a factor $e^{2\pi}$ in the Schwarzschild case \cite{Sneppen2021},
while Wielgus discussed photon rings as a test of non-Kerr metrics
\cite{Wielgus2021}. The use of such observables to constrain spacetime geometry
has been discussed by Walia, Kocherlakota, Chang and Salehi \cite{Walia2025}
and by Urso, Vincent, Wielgus, Paumard and Perrin \cite{Urso2025}, and, in the
context of EHT data analysis, by Vagnozzi and collaborators \cite{Vagnozzi2023}
and Kocherlakota and collaborators \cite{Kocherlakota2021}. Related work on the
topology of light rings is due to Cunha and Herdeiro
\cite{CunhaHerdeiro2018,CunhaHerdeiro2020}.

A feature common to essentially all of these treatments is that the analytic
content resides in the trajectory. The position of the photon at an intermediate
point of its path is written explicitly, and the construction of an image
reduces to matching the endpoints of the trajectories against the source. This
is conceptually clear, but it retains a ray-by-ray structure: each ray has to be
traced individually and its endpoint tested separately. Analyses based on the
strong-deflection limit replace the exact trajectory by an expansion valid only
near the critical impact parameter, whereas parametrisations such as that of
Gralla and Lupsasca, although exact, are used mainly to generate trajectories
numerically. Related work has also sought analytic or algebraic descriptions of
the map between a source and its image in the Schwarzschild spacetime
\cite{Zhu2025,Wu2025}.

In this paper we follow a different route. Instead of treating the solution of
the trajectory as an intermediate step, we take the relation between the source
and its image to be the analytic object of interest. We construct the mapping
defined between the apparent-plane coordinates $(R,\Theta)$ and the source
coordinates $(r,\theta,\phi)$, and refer to it as the photograph function. The
branch structure of this function corresponds to the number of times a ray winds
around the black hole; its critical behaviour is governed by the photon sphere
and the shadow boundary; and the intermediate trajectory coordinates are never
needed.

The construction relies on the spherical symmetry of the Schwarzschild spacetime
and on its conserved quantities. For a scattered ray incident from infinity the
orbit is characterised by the energy $E$ and the angular momentum $L$. Once the
apparent-plane coordinates are fixed, the total deflection of a ray that passes
through its perihelion and returns to infinity is given by a single integral
which, in the case of a source at infinity, can be written analytically as a
combination $2z_p$ of complete and incomplete elliptic integrals of the first
kind. Since a scattered ray is symmetric about its perihelion, the direction of
the source is then fixed entirely by $2z_p$ and by the apparent-plane
coordinates. The photograph function therefore takes the form of an explicit
multivalued map whose branches are labelled by the interval in which $2z_p$
lies. Outside the shadow, $R>3\sqrt{3}M$, this map gives the celestial direction
corresponding to any image position without solving the intermediate null
geodesic. After a suitable radial stretching of the image plane, the two-layer
self-similar structure formed by the higher-order images can be displayed with
high precision. When the source lies in the equatorial plane the situation
differs: for the first image, rays that cross the shadow region are absorbed,
while scattered rays may reach the source before reaching the perihelion.
Treating these cases separately again yields a photograph function.

Compared with existing analytic approaches, the contribution of this paper is
not a new solution of the trajectory but a change in the level at which the
analysis is performed: from the analyticity of the trajectory to the analyticity
of the endpoint mapping. As a result the photograph can be computed directly
from the conserved quantities and the deflection angle, without evaluating the
trajectory integral. In the Schwarzschild case the framework is complete. Its
extension to the Kerr spacetime requires the introduction of the Carter constant
and a treatment of the fact that the orbit is no longer symmetric about a
perihelion, and is left for future work.

The paper is organised as follows. Section~\ref{sec:setup} introduces the two
coordinate systems and the equations of null geodesics.
Section~\ref{sec:function} discusses the analytic solution of these equations
and sets out the construction of the photograph function.
Section~\ref{sec:infinity} treats sources at infinity and gives the explicit
function together with the classification of its branches and the self-similar
structure of the image. Section~\ref{sec:equator} treats sources in the
equatorial plane, corresponding to the photograph of a thin accretion disk.
Section~\ref{sec:conclusions} summarises the results and discusses possible
extensions. 

% ===========================================================================
\section{Basic setup}
\label{sec:setup}
% ===========================================================================

\begin{figure}[!htbp]
\centering
% ---- Left: zoomed-in detail image ----
\begin{subfigure}{0.24\linewidth}
    \centering
    \begin{tikzpicture}[remember picture, boximg]
        % Insert the zoomed image (replace with actual fig2.png)
        \node[anchor=south west, inner sep=0] (img1) {\includegraphics[width=\dimexpr\linewidth-1pt\relax]{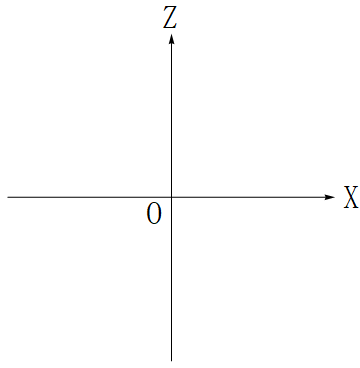}};
        % Optional label
        \node[red, font=\small] at (0.5,0.5){};
        \draw (img1.south west) rectangle (img1.north east);
    \end{tikzpicture}
    % No subcaption
\end{subfigure}
\hfill
% ---- Right: full image with red box ----
\begin{subfigure}{0.64\linewidth}
    \centering
    \begin{tikzpicture}[remember picture, boximg]
        % Insert original image
        \node[anchor=south west, inner sep=0] (img) {\includegraphics[width=\linewidth]{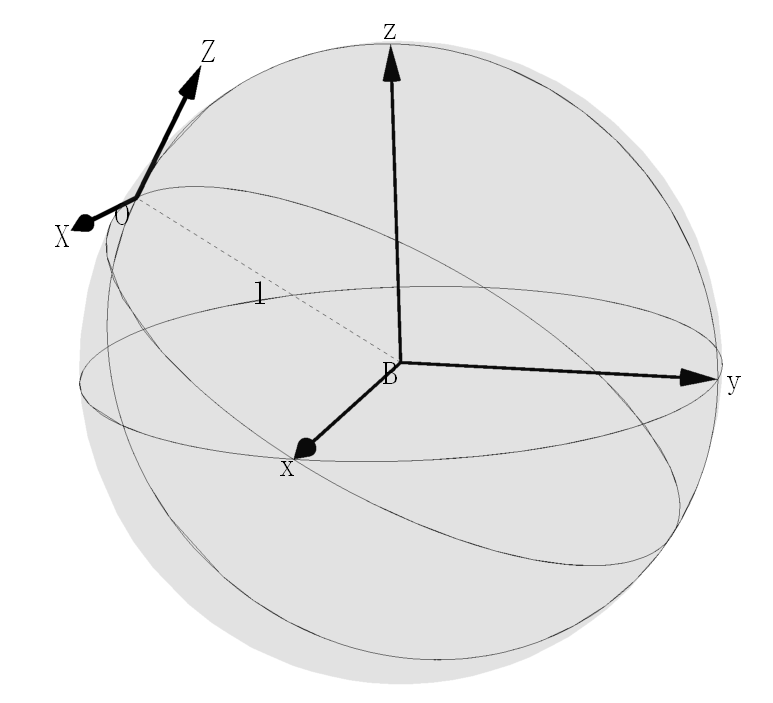}};
        % Relative coordinate system (0~1)
        \begin{scope}[x={(img.south east)}, y={(img.north west)}]
            % Red dashed rectangle indicating the zoomed region
            \draw[red, thick, dashed] (0.07,0.62) rectangle (0.28,0.95);
            % Coordinate for connection start point (lower right corner of the box)
            \coordinate (B1) at (0.28,0.62);
            % Label
            \node[red, font=\small] at (0.38,0.62) {};
        \end{scope}
    \end{tikzpicture}
    % No subcaption
\end{subfigure}

% ---- Connection line from the box (right) to the zoomed image (left) ----
\begin{tikzpicture}[overlay,boximg]
   		\draw (B1) -- (img1);
\end{tikzpicture}

\caption{The celestial coordinate and the apparent coordinate.}
\label{fig:coordinates}
\end{figure}

We introduce two coordinate systems. The first describes the position of the
source and is referred to as the celestial coordinate; the second describes the
position of the image and is referred to as the apparent coordinate. As shown in
Fig.~\ref{fig:coordinates}, the celestial coordinate system is a
three-dimensional system (Cartesian or spherical) centred on the Schwarzschild
black hole $B$. The apparent coordinate system is a two-dimensional plane
centred on the observer $O$; it is embedded in the sphere of radius $l$ centred
on the black hole and is tangent to that sphere. Here $l$ is the distance from
the observer to the centre of the black hole. We take the origin of the apparent
coordinate to lie on the negative $y$ axis, with $x\parallel X$, and denote by
$\Psi$ the angle between $OB$ and $Bz$. The black hole itself does not emit
light, so a photograph of a black hole shows the other light sources present in
the celestial coordinate. Between an image point in the apparent coordinate and
a source point in the celestial coordinate there is a null geodesic.

After an appropriate calculation, described in Appendix~\ref{app:geodesics},
the equations of null geodesics in the Schwarzschild spacetime can be written as
a system of first-order ordinary differential equations,
\begin{align}
\left(\frac{du}{d\lambda}\right)^{2}
  &= u^{4}R^{2}f(u), \label{eq:eom-u} \\[2pt]
\left(\frac{d\theta}{d\lambda}\right)^{2}
  &= u^{4}R^{2}\left(1-\csc^{2}\theta\,\sin^{2}\Theta\,\sin^{2}\Psi\right),
  \label{eq:eom-theta} \\[2pt]
\left(\frac{d\phi}{d\lambda}\right)^{2}
  &= u^{4}R^{2}\csc^{4}\theta\,\sin^{2}\Theta\,\sin^{2}\Psi,
  \label{eq:eom-phi}
\end{align}
where
\begin{equation}
f(u)=2Mu^{3}-u^{2}
+\frac{1}{R^{2}}\left(1-\frac{2MR^{2}}{(l^{2}+R^{2})^{3/2}}\right),
\label{eq:f}
\end{equation}
and the conserved quantities are
\begin{align}
E &= \sqrt{1-\frac{2MR^{2}}{(l^{2}+R^{2})^{3/2}}}, \label{eq:E} \\
L &= R, \label{eq:L} \\
P_{\phi} &= R\,\sin\Theta\,\sin\Psi. \label{eq:Pphi}
\end{align}
The initial conditions are
\begin{equation}
u_{i}=\frac{1}{l},\qquad \theta_{i}=\Psi,\qquad \phi_{i}=-\frac{\pi}{2}.
\label{eq:ic}
\end{equation}
We are not concerned with the evolution of the time coordinate $t$, and the
equation for $dt/d\lambda$ has therefore been omitted. Since actual black holes
are very far from us, for convenience we usually take $l\to\infty$.

For the framework described above to be valid, the image plane must be much
smaller than the sphere on which the observer sits. It should be noted that the
distances in the apparent coordinate are not physical distances. Taking
Sagittarius A*, whose mass is $M_{A^{*}}=4.31\times10^{6}M_{\odot}$, one obtains
a shadow radius
\begin{equation}
R_{s}=\frac{3\sqrt{3}\,GM_{A^{*}}}{c^{2}}=3.3\times10^{10}\,\mathrm{m},
\end{equation}
which does not correspond to the actual radius of the image plane. In fact,
parallel light must be focused onto the image plane by a lens, so that a
magnification factor exists between the physical object and its image. The image
plane may therefore be chosen arbitrarily small.

% ===========================================================================
\section{Null geodesics and the photograph function}
\label{sec:function}
% ===========================================================================

Equations~\eqref{eq:eom-u}--\eqref{eq:eom-phi} admit analytic solutions. These
solutions differ slightly according to the type of the geodesic. For rays
incident from infinity the orbits fall into three classes: rays absorbed by the
black hole; rays scattered back to infinity; and rays that approach the photon
sphere. The third class is critical and lies between the other two; it satisfies
$f(u)=0$ together with $f'(u)=0$. From these conditions one obtains the photon
sphere radius $r_{c}=1/u_{c}=3M$ and the shadow radius $R_{c}=3\sqrt{3}M$. For
simplicity we consider only the scattering solutions. For the scattering case,
$u\in(0,1/3M)$ and $R>3\sqrt{3}M$, the analytic solutions are
\begin{equation}
u(z)=\frac{-\frac{1}{R}\wp'(z)+\frac{M}{2R^{2}}}
            {2\left(\wp(z)+\frac{1}{12}\right)},
\label{eq:uofz}
\end{equation}
\begin{equation}
\begin{aligned}
\pm z={}&\arcsin\!\frac{\sin\Psi}{\sqrt{1-\sin^{2}\Theta\sin^{2}\Psi}}\\
      &-\arcsin\!\frac{\sin\theta}{\sqrt{1-\sin^{2}\Theta\sin^{2}\Psi}},
\end{aligned}
\label{eq:zoftheta}
\end{equation}
\begin{equation}
\begin{aligned}
\pm\left(\phi+\frac{\pi}{2}\right)
={}& \arctan\!\frac{\sin\Theta\sin\Psi\cos\Psi}
        {\sqrt{1-\sin^{2}\Theta\sin^{2}\Psi-\cos^{2}\Psi}} \\
 &{}-\arctan\!\frac{\sin\Theta\sin\Psi\cos\theta}
        {\sqrt{1-\sin^{2}\Theta\sin^{2}\Psi-\cos^{2}\theta}},
\end{aligned}
\label{eq:phioftheta}
\end{equation}
where $\wp(z;g_{2},g_{3})$ is the Weierstrass elliptic function satisfying
$(\wp')^{2}=4\wp^{3}-g_{2}\wp-g_{3}$, with invariants
\begin{equation}
g_{2}=\frac{1}{12},\qquad
g_{3}=\frac{1}{216}-\frac{M^{2}}{4R^{2}} .
\label{eq:invariants-main}
\end{equation}
The function $z$ is defined by
\begin{equation}
z=\int_{0}^{u}\frac{du'}{\sqrt{f(u')}} .
\label{eq:zdef}
\end{equation}
The derivation is given in Appendix~\ref{app:solutions}.

Once a source is present the situation becomes more involved. The usual approach
is ray tracing, whose guiding idea is the following: light is propagated
backwards from the image plane, and a point is illuminated if the corresponding
ray can be connected to a source, while it remains dark otherwise. This
procedure is workable but cumbersome, since every ray has to be checked
individually against the source. It is natural to ask whether the image position
can be obtained directly from the source position, or conversely, without
solving the null geodesics. In this connection we note that
Eqs.~\eqref{eq:uofz}--\eqref{eq:phioftheta} describe a family of rays
that reaches every point $(R,\Theta)$ of the apparent coordinate, while the
source is the other endpoint $(r,\theta,\phi)$ of the null geodesic. Once one
endpoint is specified, the relations~\eqref{eq:uofz}--\eqref{eq:phioftheta}
already constrain the other. We refer to the direct functional relation between
the two endpoints as the photograph function, and show below that such an
end-to-end relation can indeed be constructed.

% ===========================================================================
\section{Sources at infinity}
\label{sec:infinity}
% ===========================================================================

As a simple and practical example we consider all sources to lie on the
celestial sphere at infinity centred on the black hole $B$. Since the
Schwarzschild spacetime is spherically symmetric, we may place the observer $O$
on the negative $y$ axis without loss of generality, that is, $\Psi=\pi/2$.
Consider a scattered photon, which travels from infinity (the observer) along
its geodesic to the point closest to the black hole, the perihelion
$r_{p}=1/u_{p}$, and then travels back to infinity along the same geodesic. In
what follows we examine the trajectory in order to construct directly the
analytic function relating the source to the image point.

\subsection{Ray trajectory}
\label{sec:infinity-trajectory}

\begin{figure}[!htbp]
\centering
\includegraphics[width=0.8\linewidth]{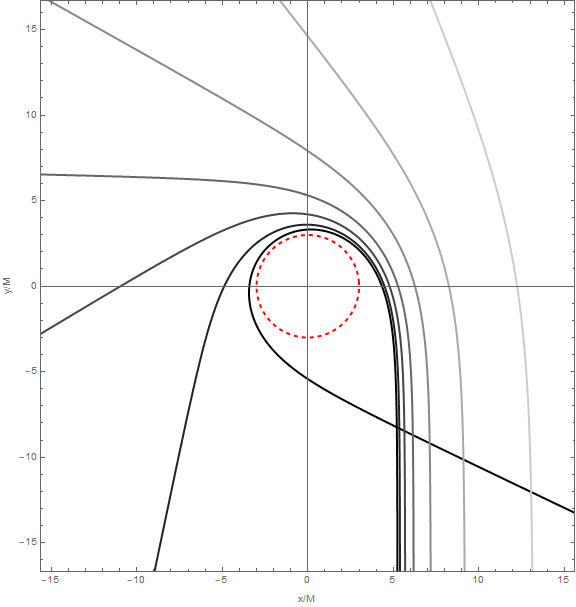}
\caption{The null geodesic constraint in the equatorial plane. The dashed line
denotes the photon sphere $r_{\mathrm{ph}}=3M$.}
\label{fig:geodesic}
\end{figure}

For simplicity we consider rays that propagate in the equatorial plane only.
This means $\theta=\pi/2$ and $\Theta=\pm\pi/2$. Rays with other values of
$\Theta$ follow from spherical symmetry by an appropriate rotation. Since the
motion is confined to a two-dimensional plane, the exact
solution~\eqref{eq:uofz} reduces to
\begin{equation}
u(\phi)=\frac{-\frac{1}{R}\wp'\!\left(\phi+\frac{\pi}{2}\right)
             +\frac{M}{2R^{2}}}
            {2\left(\wp\!\left(\phi+\frac{\pi}{2}\right)+\frac{1}{12}\right)^2},
\label{eq:uofphi}
\end{equation}
whose numerical solution is shown in Fig.~\ref{fig:geodesic}.

For the discussion of the end-to-end function such a detailed treatment is not
required. Since both the image point and the source are at infinity, it suffices
to determine the total deflection angle between them. The deflection angle of a
ray travelling from the image point to the perihelion is
\begin{equation}
z_{p}=\int_{0}^{u_{p}}\frac{du'}{\sqrt{f(u')}}
     =\frac{2}{\sqrt{2M(e_{3}-e_{1})}}
       \Big[K(k)-F(\theta_{0}\,|\,k)\Big],
\label{eq:zp}
\end{equation}
where $K(k)$ is the complete elliptic integral of the first kind and
$F(\theta_{0}|k)$ the incomplete one, with
\begin{equation}
k=\sqrt{\frac{e_{2}-e_{1}}{e_{3}-e_{1}}},\qquad
\theta_{0}=\arcsin\sqrt{\frac{-e_{1}}{e_{2}-e_{1}}} .
\label{eq:modulus}
\end{equation}
Here $e_{1}<0<e_{2}<e_{3}$ are the three real roots of $f(u)=0$, whose
expressions can be written as
\begin{equation}
\begin{aligned}
e_{k}&=\frac{1}{6M}\left[1+2\cos\left(\alpha+\frac{2\pi(k-3)}{3}\right)\right],\\
\alpha&=\arccos\!\left(1-\frac{54M^{2}}{R^{2}}\right)\in\left(0,\frac{\pi}{3}\right).
\end{aligned}
\label{eq:roots}
\end{equation}
The perihelion is the second zero, $u_{p}=e_{2}$. For a complete scattered ray,
shown in Fig.~\ref{fig:geodesic}, the part between the image point and the
perihelion and the part between the perihelion and the source are symmetric. The
total deflection angle is therefore twice the deflection of the incoming (or
outgoing) part, that is, $2z_{p}$. Once $2z_{p}$ is known, the source
coordinates can be obtained directly from the image coordinates, without solving
the null geodesics.

\subsection{The photograph function}
\label{sec:infinity-function}

\begin{figure*}[!htbp]
\centering
\setlength{\tabcolsep}{10pt}
  \renewcommand{\arraystretch}{1.0}
  \begin{tabular}{cc} 
    \includegraphics[width=0.35\textwidth]{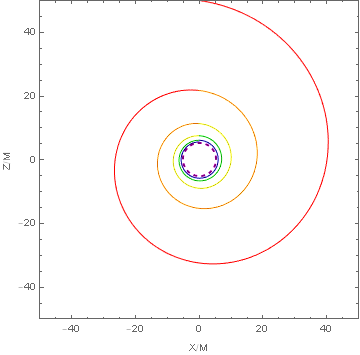} &
    \includegraphics[width=0.35\textwidth]{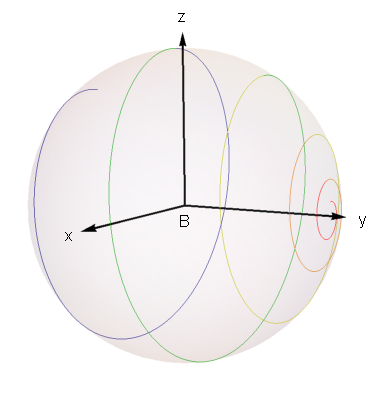} 
  \end{tabular}
\caption{The image (left) and the light source (right) on the celestial sphere.
The celestial sphere is drawn as a unit sphere for illustration, while its true
radius is $l\to\infty$. The dashed line in the left panel denotes the shadow
radius $R_{s}=3\sqrt{3}M$. Two points of the same colour in the apparent and
celestial coordinates are linked by a null geodesic which need not be computed
explicitly.}
\label{fig:correspondence}
\end{figure*}

Consider a point $(R,\Theta)$ in the apparent plane. The ray arriving at this
point, traced backwards, reaches the source $(r,\theta,\phi)$ on the celestial
sphere. Since the source is assumed to be at infinity, $r\to\infty$, it is
sufficient to describe its position by the unit vector
$(n_{x},n_{y},n_{z})$. An appropriate calculation gives
\begin{equation}
\begin{pmatrix} n_{x} \\ n_{y} \\ n_{z} \end{pmatrix}
=\begin{pmatrix} \sin(2z_{p})\,\sin\Theta \\[2pt]
                -\cos(2z_{p}) \\[2pt]
                \sin(2z_{p})\,\cos\Theta \end{pmatrix}.
\label{eq:photo-infinity}
\end{equation}
After removing the shadow region $R\leq3\sqrt{3}M$ of the photograph, the
relation above gives the correspondence between the image position $(R,\Theta)$
and the source position $(n_{x},n_{y},n_{z})$ directly, without solving the
intermediate null geodesics. Figure~\ref{fig:correspondence} shows one such
correspondence.

More often one wishes to determine the image position $(R,\Theta)$ from a known
source position $(n_{x},n_{y},n_{z})$, which requires inverting
Eq.~\eqref{eq:photo-infinity}. Apart from certain special points the inverse
function falls into two classes, listed in Table~\ref{tab:two-classes}. In
particular, when $n_{y}^{2}=1$ one has $\sin 2z_{p}=0$, and the angle $\Theta$
may take any value in $[0,2\pi]$. The point is then mapped to a ring, which is
the Einstein ring. This special case is listed separately in
Table~\ref{tab:einstein}.

\begin{table*}[!t]
\caption{Two classes of solutions for $n_{y}^{2}\neq1$, with
$k=1,2,3,4,\dots$. $R_{i}$ denotes the solution of
$z_{p}(R_{i})=\pm\frac{1}{2}\arccos(-n_{y})+i\pi$.}
\label{tab:two-classes}
\begin{ruledtabular}
\begin{tabular}{lll}
Type I & $z_{p}=-\frac{1}{2}\arccos(-n_{y})+k\pi$,
         $\Theta=\arctan\frac{n_{x}}{n_{z}}+\pi$,&
         $(X,Z)=\left(-\frac{R_{i}n_{x}}{\sqrt{1-n_{y}^{2}}},\,
         -\frac{R_{i}n_{z}}{\sqrt{1-n_{y}^{2}}}\right)$ \\[6pt]
Type II & $z_{p}=\frac{1}{2}\arccos(-n_{y})+k\pi$,
          $\Theta=\arctan\frac{n_{x}}{n_{z}}$,&
          $(X,Z)=\left(\frac{R_{i}n_{x}}{\sqrt{1-n_{y}^{2}}},\,
          \frac{R_{i}n_{z}}{\sqrt{1-n_{y}^{2}}}\right)$ \\
\end{tabular}
\end{ruledtabular}
\end{table*}

\begin{table*}[!t]
\caption{Two classes of solutions for $n_{y}^{2}=1$.}
\label{tab:einstein}
\begin{ruledtabular}
\begin{tabular}{lll}
Type I & $z_{p}=-\frac{1}{2}\arccos(-n_{y})+k\pi$,
         $\Theta\in[0,2\pi]$,& $R_{i}=z_{p}^{-1}(z_{i})$ \\[6pt]
Type II & $z_{p}=\frac{1}{2}\arccos(-n_{y})+k\pi$,
          $\Theta\in[0,2\pi]$, &$R_{i}=z_{p}^{-1}(z_{i})$ \\
\end{tabular}
\end{ruledtabular}
\end{table*}

\begin{table*}[!t]
\caption{Some special apparent coordinates.}
\label{tab:coordinate}
\begin{ruledtabular}
\begin{tabular}{llll}
 &  North Pole  & South Pole &  Einstein Ring\\[6pt]
$k=1$ & $(0,-6.16757M)$ & $(0,6.16757M)$ & $R=5.35696M$ \\[6pt]
$k=2$ & $(0,5.22794M)$ & $(0,-5.22794M)$ & $R=5.20268M$  \\[6pt]
$k=3$ & $(0,-5.19752M)$ & $(0,5.19752M)$ & $R=5.196433M$ \\          
\end{tabular}
\end{ruledtabular}
\end{table*}

\begin{figure*}[!htbp]
\centering
\setlength{\tabcolsep}{6pt}
  \renewcommand{\arraystretch}{1.0}
  \begin{tabular}{ccc}
    \includegraphics[width=0.28\textwidth]{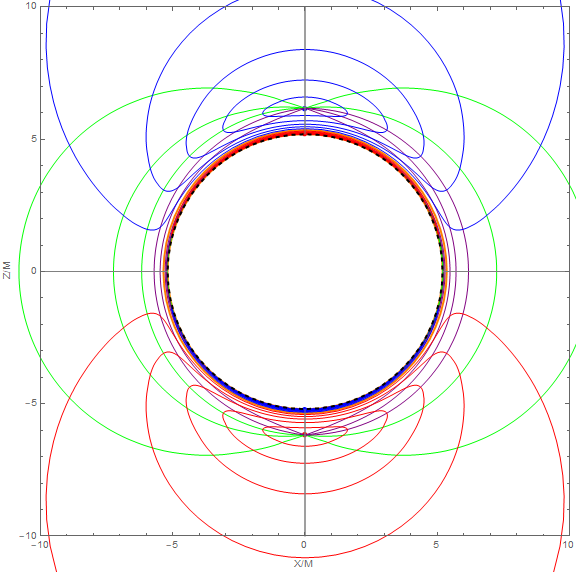} &\includegraphics[width=0.28\textwidth]{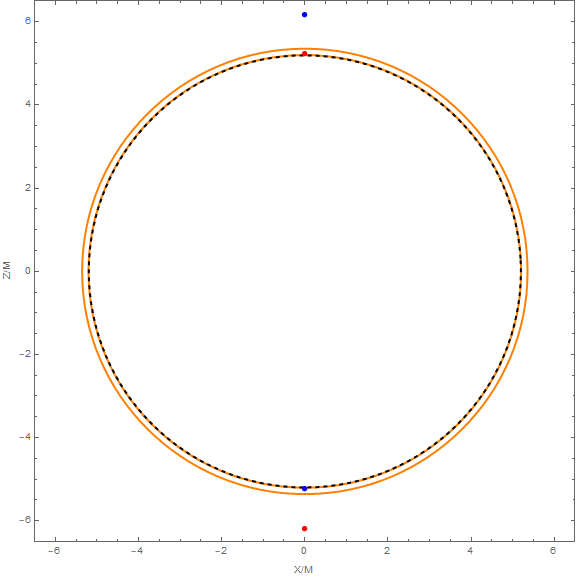} &
    \includegraphics[width=0.28\textwidth]{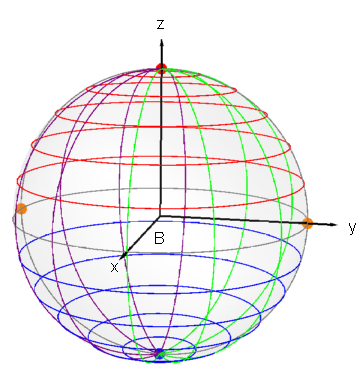}
  \end{tabular}
\caption{The image (left), the Einstein ring (middle) and the
latitude--longitude grid (right).}
\label{fig:grid}
\end{figure*}

In addition, the function $z_{p}(R)$ has to be inverted. A mathematical analysis
shows that $z_{p}(R)$ is one-to-one in its argument $R$, so that its inverse
exists and is unique but cannot be written in a simple analytic form; a simple
bisection method is sufficient to construct the inverse numerically. Some special apparent coordinates are given in Table~\ref{tab:coordinate}. 
Using the relations above, the photograph of a latitude--longitude grid of sources is
obtained as shown in Fig.~\ref{fig:grid}.

\begin{figure*}[!htbp]
\centering
\setlength{\tabcolsep}{10pt}
  \renewcommand{\arraystretch}{1.0}
  \begin{tabular}{cc} 
    \includegraphics[width=0.35\textwidth]{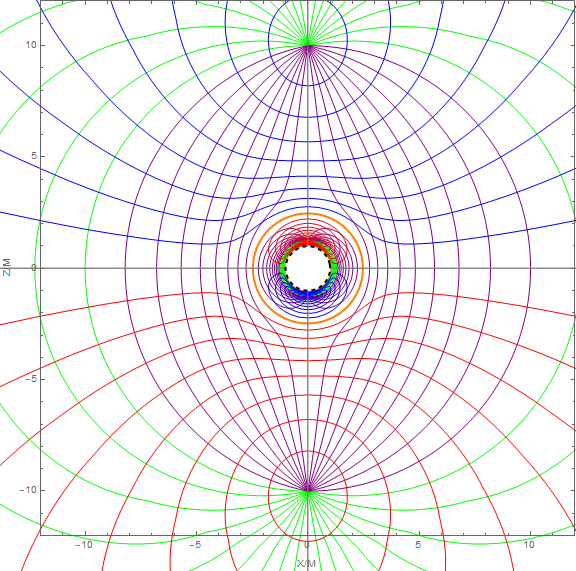} &
    \includegraphics[width=0.35\textwidth]{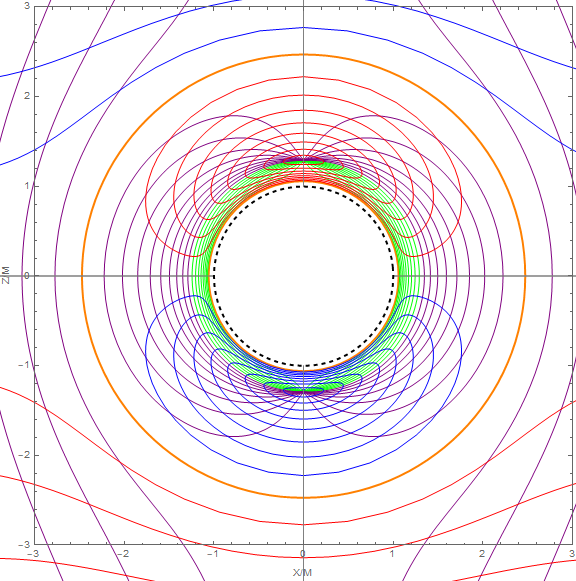} 
  \end{tabular}
\caption{The structure of the image (shown for $k=1,2$). A suitable radial
stretching has been applied to the image plane in order to make the fine
structure visible: the shadow radius is mapped to a radius of $1$, while the
radial coordinate of the first image of the north pole is stretched to $10$.}
\label{fig:structure}
\end{figure*}

When the photograph is stretched radially in an appropriate way, its internal
structure becomes clearer, as shown in Fig.~\ref{fig:structure}. One finds a
two-layer structure that is inverted with respect to itself and repeats without
limit as the shadow radius $R=3\sqrt{3}M$ is approached. Previously this
structure could only be obtained by solving a large number of null geodesics;
it can now be displayed directly, quickly and with high precision.

% ===========================================================================
\section{Sources in the equatorial plane}
\label{sec:equator}
% ===========================================================================

When an accretion disk surrounds the black hole, the accretion flow heats the
disk and causes it to radiate. This radiation can act as a light source and
illuminate the black hole. Such sources are generally assumed to lie in the
equatorial plane; their photograph depends on the position $\Psi$ of the
observer, and we assume $\Psi\in(0,\pi/2)$. Since the source is confined to the
equatorial plane, the source coordinate satisfies $\theta=\pi/2$, and only the
two coordinates $r$ and $\phi$ remain to be determined. The appearance of the
image is related to the number of times the light winds around the black hole.

\subsection{Ray trajectory}
\label{sec:equator-trajectory}

As an example we consider the ray trajectory in the $xy$ plane, for which
$\phi=\pm\pi/2$ and $\sin\Theta=0$. For simplicity we consider the imaging
process in which light reaches the apparent coordinate along the shortest path,
that is, the first image. Suppose that light propagates backwards from the image
plane towards the source. There are two possible paths.

\emph{(1) The angle $\theta\in[0,\pi/2]$ increases monotonically.} In this case
$\cos\Theta=-1$, and
\begin{equation}
z=\int_{\Psi}^{\theta}d\theta=\theta-\Psi .
\label{eq:path1}
\end{equation}

\emph{(2) The angle $\theta\in[0,\pi/2]$ first decreases to $0$ and then
increases.} In this case $\cos\Theta=1$, and
\begin{equation}
z=
\begin{cases}
\displaystyle-\int_{\Psi}^{\theta}d\theta=\Psi-\theta, & \text{first segment},\\[8pt]
\displaystyle\left(-\int_{\Psi}^{0}+\int_{0}^{\theta}\right)d\theta=\Psi+\theta,
 & \text{second segment}.
\end{cases}
\label{eq:path2}
\end{equation}

\begin{figure}[!htbp]
\centering
\includegraphics[width=0.8\linewidth]{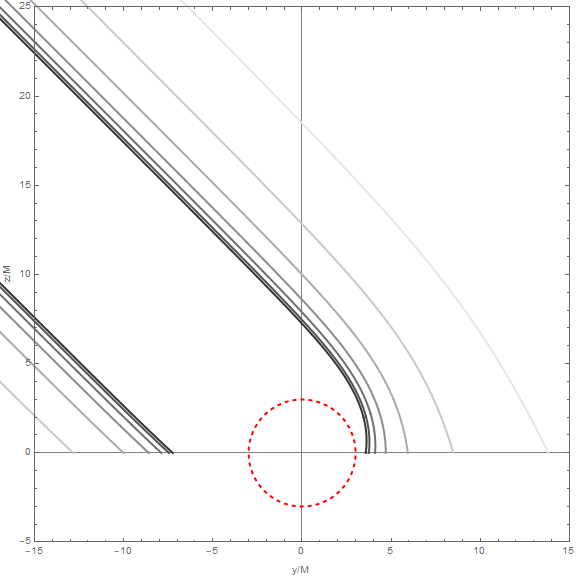}
\caption{The two classes of trajectories by which a source reaches the
equatorial plane, for $\Psi=\pi/4$. The dashed line marks the photon sphere.}
\label{fig:trajectories}
\end{figure}

Taking the change in $\theta$ as the step and substituting into
Eq.~\eqref{eq:uofz}, one obtains the value of $u$ at each step. Plotting these
points gives the ray trajectories shown in Fig.~\ref{fig:trajectories}. In fact,
for a photograph of a black hole only the functional relation between the
starting and ending points of the rays is required; it is not necessary to
compute every point on the geodesic.

\subsection{The photograph function}
\label{sec:equator-function}

As in the preceding section, we may construct directly the photograph function
relating the image point $(R,\Theta)$ and the source $(r,\pi/2,\phi)$,
\begin{equation}
\begin{pmatrix} n_{x} \\ n_{y} \\ n_{z} \end{pmatrix}
=\begin{pmatrix} \sin\Theta\,\sin z \\[2pt]
                \cos\Theta\,\frac{\sin z}{\cos\Psi} \\[2pt]
                0 \end{pmatrix},
\label{eq:photo-equator}
\end{equation}
where $n_{x}=\cos\phi$ and $n_{y}=\sin\phi$.
According to the analysis above, the null geodesics traced backwards from the
photograph fall into two classes.

\emph{(1) Positive $Z$ axis, $\Theta\in(-\pi/2,\pi/2)$.} A ray emitted from this
region has $\theta$ decreasing first from $\Psi$ to $\theta_{\min}$ and then
increasing to $\pi/2$. Here $\theta_{\min}$ is determined by the vanishing of
the denominator $1-\csc^{2}\theta\sin^{2}\Theta\sin^{2}\Psi$. One then has
\begin{equation}
\begin{aligned}
z_{s1}&=\left(-\int_{\theta_{i}=\Psi}^{\theta_{\min}}
            +\int_{\theta_{\min}}^{\pi/2}\right)
        \frac{d\theta}{\sqrt{1-\csc^{2}\theta\sin^{2}\Theta\sin^{2}\Psi}}\\
     &=\pi-\arcsin\!\frac{\cos\Psi}{\sqrt{1-\sin^{2}\Theta\sin^{2}\Psi}} .
\end{aligned}
\label{eq:zs1}
\end{equation}

\begin{figure*}[!htbp]
   \centering
    \begin{subfigure}[b]{0.19\textwidth}
        \centering
        \includegraphics[width=\textwidth]{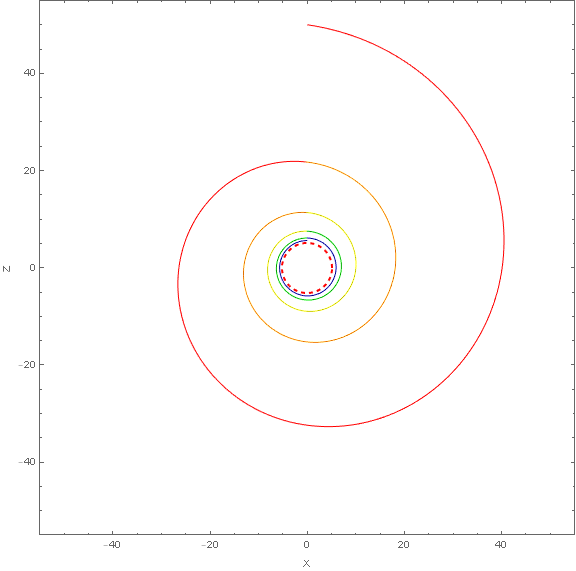}
        \caption{Direct image}
        \label{fig:sub1}
    \end{subfigure}
    \hfill
    \begin{subfigure}[b]{0.19\textwidth}
        \centering
        \includegraphics[width=\textwidth]{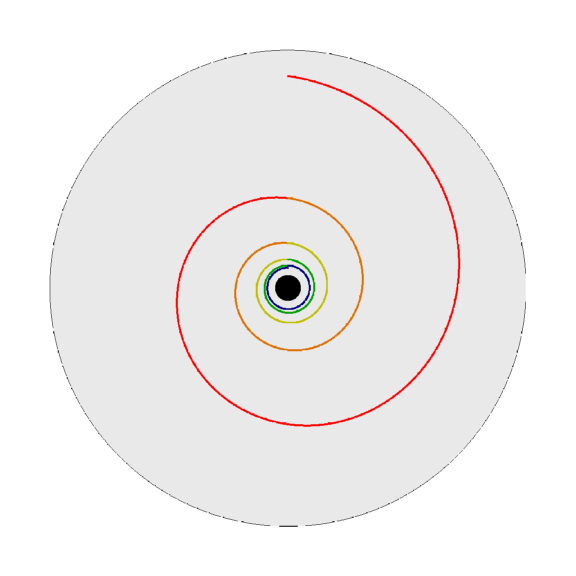}
        \caption{$\Psi=0.01$}
        \label{fig:sub2}
    \end{subfigure}
    \hfill
    \begin{subfigure}[b]{0.19\textwidth}
        \centering
        \includegraphics[width=\textwidth]{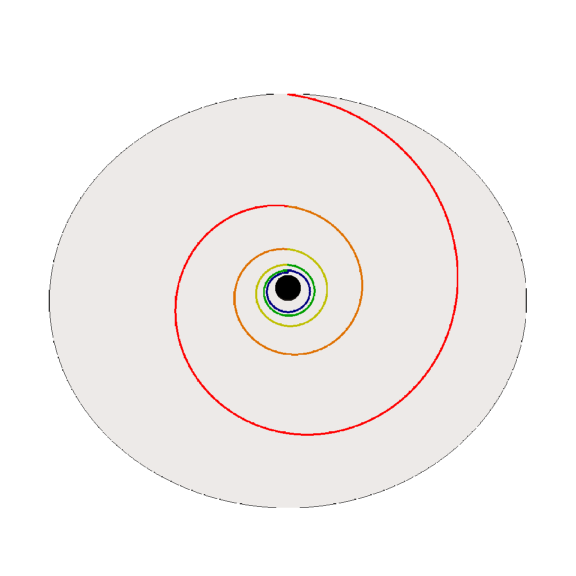}
        \caption{$\Psi=\pi/6$}
        \label{fig:sub3}
    \end{subfigure}
    \hfill
    \begin{subfigure}[b]{0.19\textwidth}
        \centering
        \includegraphics[width=\textwidth]{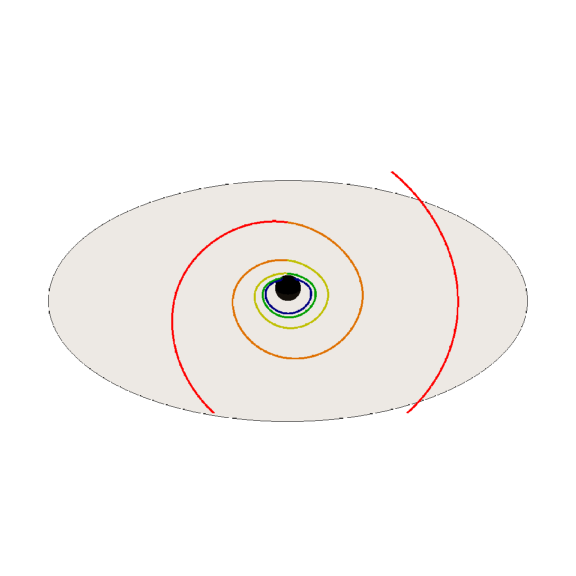}
        \caption{$\Psi=\pi/3$}
        \label{fig:sub4}
    \end{subfigure}
    \hfill
    \begin{subfigure}[b]{0.19\textwidth}
        \centering
        \includegraphics[width=\textwidth]{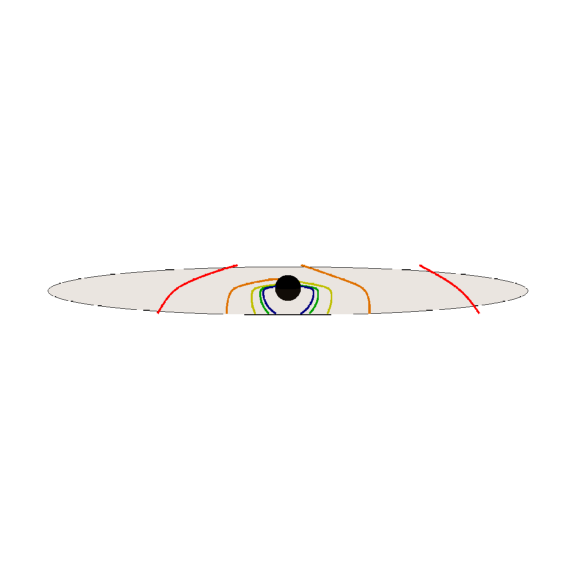}
        \caption{$\Psi=\pi/2-0.1$}
        \label{fig:sub5}
    \end{subfigure}
\caption{The image (leftmost panel) compared with the light source for
different values of $\Psi$. Panels (b)--(e) show the three-dimensional
configuration for an observer viewing the equatorial plane from different
angles $\Psi$ at a distance $l=10M$. The black sphere in the middle represents
the photon sphere.}
\label{fig:equator-correspondence}
\end{figure*}

\emph{(2) Negative $Z$ axis, $\Theta\in(\pi/2,3\pi/2)$.} A ray emitted from this
region has $\theta$ increasing monotonically from $\Psi$ to $\pi/2$. Therefore
\begin{equation}
\begin{aligned}
z_{s2}&=\int_{\theta_{i}=\Psi}^{\pi/2}
       \frac{d\theta}{\sqrt{1-\csc^{2}\theta\sin^{2}\Theta\sin^{2}\Psi}}\\
     &=\arcsin\!\frac{\cos\Psi}{\sqrt{1-\sin^{2}\Theta\sin^{2}\Psi}} .
\end{aligned}
\label{eq:zs2}
\end{equation}
After some algebra the source point is found to be
\begin{equation}
\begin{aligned}
(x,y,0)=\Bigg(&
\frac{\sin\Theta\cos\Psi}{u_{s}\sqrt{1-\sin^{2}\Theta\sin^{2}\Psi}},\\
&\frac{\cos\Theta}{u_{s}\sqrt{1-\sin^{2}\Theta\sin^{2}\Psi}},\;0\Bigg),
\end{aligned}
\label{eq:source-point}
\end{equation}
where $u_{s}$ is obtained by substituting the corresponding $z_{s1}$ or $z_{s2}$
into Eq.~\eqref{eq:uofz}. Figure~\ref{fig:equator-correspondence} shows one such
correspondence.

\begin{figure*}[!htbp]
\centering
\setlength{\tabcolsep}{4pt}
  \renewcommand{\arraystretch}{1.0}
  \begin{tabular}{cccc} 
    \includegraphics[width=0.22\textwidth]{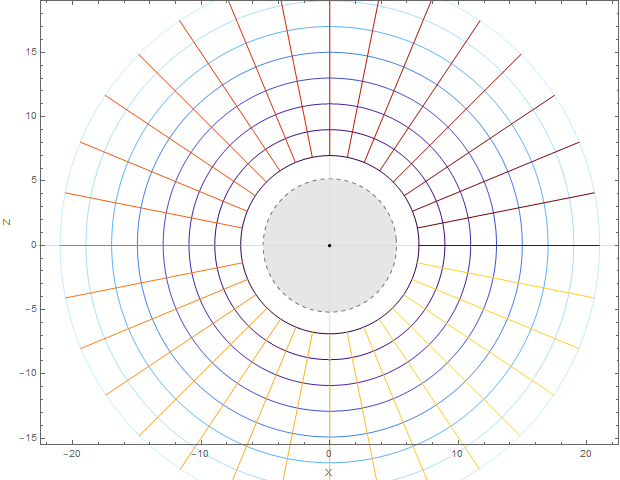} &
    \includegraphics[width=0.22\textwidth]{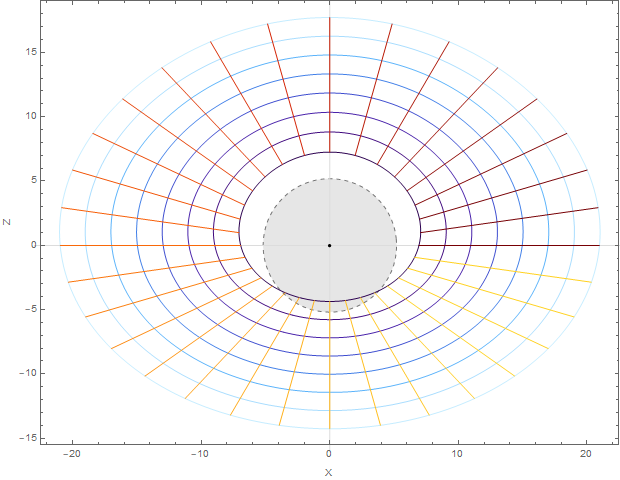} &
    \includegraphics[width=0.22\textwidth]{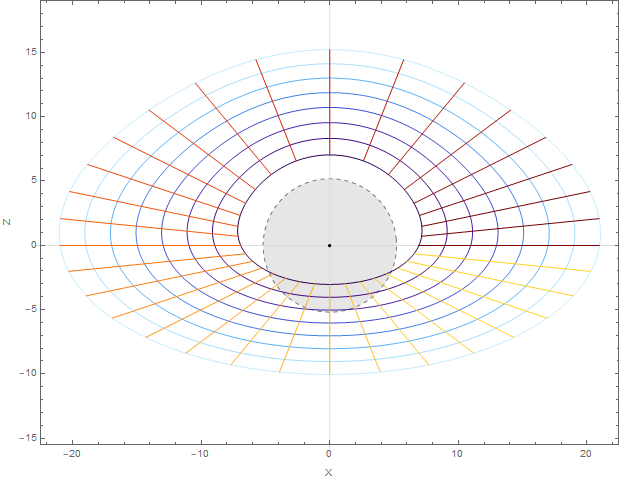} &
    \includegraphics[width=0.22\textwidth]{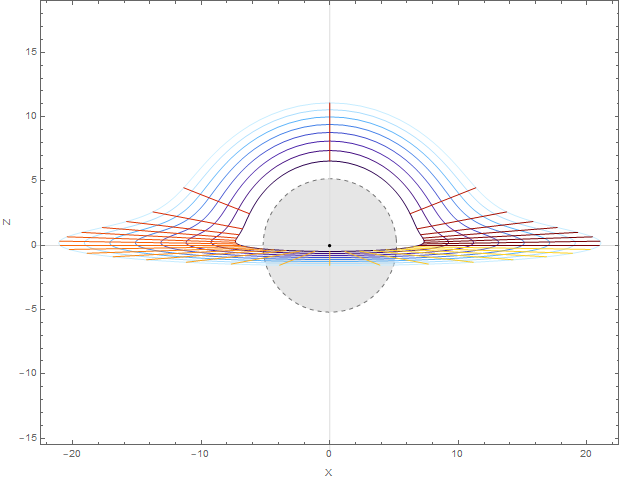} \\
      \includegraphics[width=0.22\textwidth]{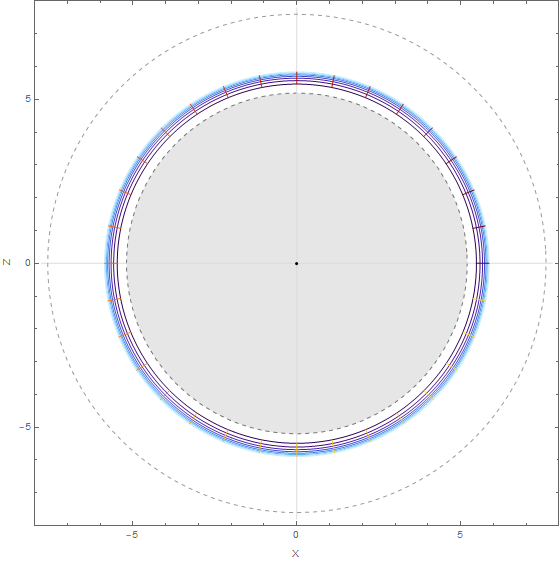} &
    \includegraphics[width=0.22\textwidth]{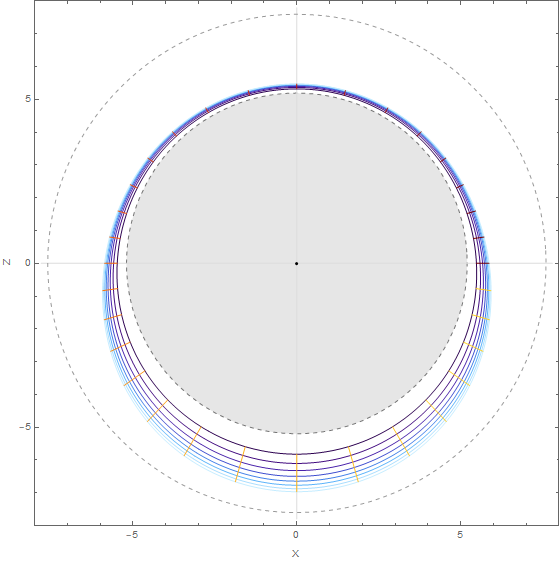} &
    \includegraphics[width=0.22\textwidth]{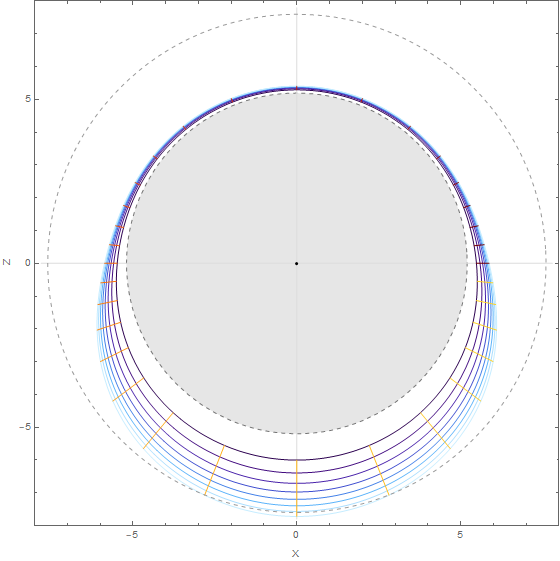} &
    \includegraphics[width=0.22\textwidth]{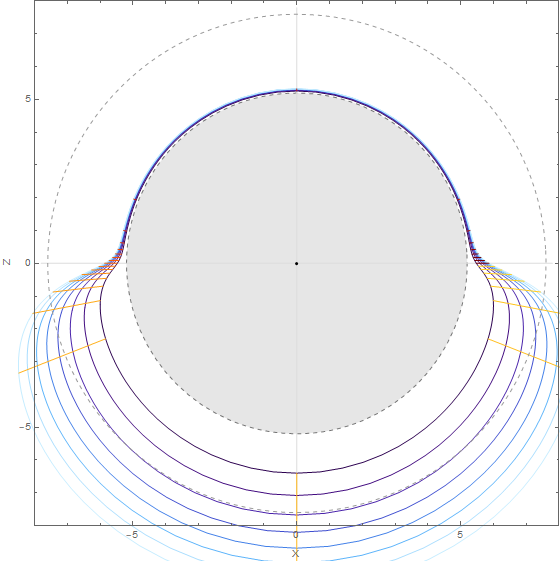} 
  \end{tabular}
\caption{The black hole image for sources lying in the equatorial plane, for
$\Psi=\pi/100$, $\pi/4$, $\pi/3$ and $19\pi/40$ from left to right. The first row shows the first imaging, and the second row shows the second imaging. The grey
region represents the black hole shadow.}
\label{fig:equator-grid}
\end{figure*}

\begin{figure}[!htbp]
\centering
\includegraphics[width=0.8\linewidth]{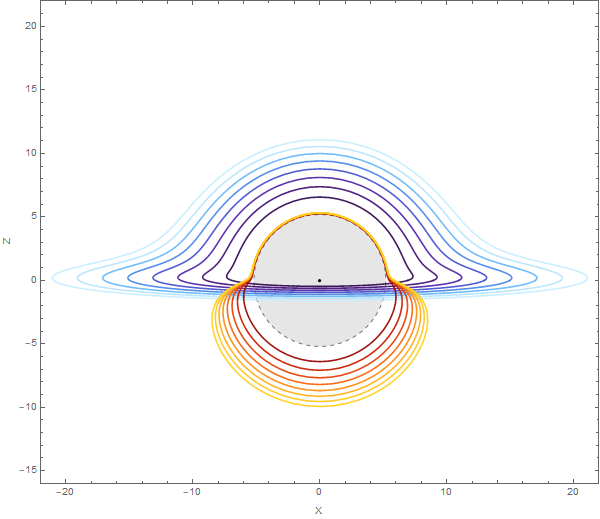}
\caption{The black hole photograph combining the first- and second-order images ($\Psi=19\pi/40$).}
\label{fig:imageJoined}
\end{figure}

Of greater interest is the converse problem: given the source position
$(n_{x},n_{y},0)=(r\cos\phi,r\sin\phi,0)$, how is the image position $(X,Z)$
obtained? For the first image one finds
\begin{equation}
\begin{aligned}
(X,Z)&=\left(
\frac{R\,n_{x}^{2}}{n_{x}^{2}+n_{y}^{2}\cos^{2}\Psi},\;
\frac{R\,n_{y}\cos\Psi}{n_{x}^{2}+n_{y}^{2}\cos^{2}\Psi}\right),\\
\Theta&=\arctan\!\frac{n_{x}}{n_{y}\cos\Psi}.
\end{aligned}
\label{eq:image-position}
\end{equation}
The result for $R$ requires a case-by-case discussion. We define the functions
\begin{align}
\begin{split}
z_{sd} &= \int_{0}^{u}\frac{dx}{\sqrt{2Mx^{3}-x^{2}+\frac{1}{27M^{2}}}}\\
       &= \ln\!\left[
          \frac{\sqrt{3}+\sqrt{1+6Mu}}{\sqrt{3}-\sqrt{1+6Mu}}\,
          \frac{\sqrt{3}-1}{\sqrt{3}+1}\right],
\end{split}
\label{eq:zsd} \\[4pt]
z_p[R_{c}] &=z_{s1}  .
\label{eq:zpRc}
\end{align}
After the corresponding algebra, the various values of $z$ are obtained as
listed in Table~\ref{tab:zcases}. Here $z$ denotes a scattered ray and $z'$ an
absorbed ray. The function $R$ is then constructed by the bisection method,
solving the equation
\begin{equation}
z=\int_{0}^{u}\frac{du'}{\sqrt{f(u',R)}}
\end{equation}
for $R$. A similar process can also be used to discuss secondary or higher imaging. Using this result, the coordinate grid on the equatorial plane near the
black hole can be mapped onto the photograph; the result is shown in
Fig.~\ref{fig:equator-grid}. Combining the two gives us the familiar image of a black hole with disk, see Fig.~\ref{fig:imageJoined}.

\begin{table*}[!t]
\caption{Solutions for $z$.}
\label{tab:zcases}
\begin{ruledtabular}
\begin{tabular}{lll}
Condition 1 & Condition 2 & $z$ \\
\hline
$\phi\in[0,\pi)$ & $R\geq R_c$ & $z=z_{s1}$ \\
                 & $R< R_c$ & $z=2z_{p}-z_{s1}$ \\[4pt]
$\phi\in[\pi,2\pi)$ & $z_{s2}\geq z_{sd}$ & $z=z_{s2}$ \\
                    & $z_{s2}<z_{sd}$ & $z'=z_{s2}$ \\
\end{tabular}
\end{ruledtabular}
\end{table*}

% ===========================================================================
\section{Conclusions and outlook}
\label{sec:conclusions}
% ===========================================================================

We have introduced two coordinate systems, one for the source and one for the
image, and have used them to establish an end-to-end functional relation for the
photograph of a black hole, which we call the photograph function. The departure
from the conventional treatment is a change in the analytic object: instead of
the trajectory of the light ray, we take the functional relation between the
source and the image point. As a unified framework for describing the photograph
of a black hole, the photograph function is comparatively direct. Explicit forms
have been given for two cases, sources on the celestial sphere at infinity and
sources in the equatorial plane.

Several limitations of the present treatment should be stated. First, the
photograph function is a map from three-dimensional to two-dimensional
coordinates, so some information is necessarily lost. It is therefore not
possible to reconstruct the three-dimensional configuration of the sources from
the two-dimensional photograph alone. Certain quantities can nevertheless be
inferred: the size of the shadow, the positions of the north and south pole
images, and the location of the Einstein ring all carry information about the
mass of the black hole. Second, the function is multivalued. This multivalued
structure is what produces the two-layer, mutually inverted and self-similar
pattern of the image, which repeats without limit as the shadow radius is
approached.

Several extensions suggest themselves. The photograph function as constructed
here describes only the positional correspondence between source and image; the
addition of a brightness relation would require the radiative transfer along the
geodesic to be expressed in the same language, which is a natural next step.
The framework is not restricted to non-luminous objects: it applies equally to
emitting bodies such as stars, and establishing the functional relation between
the surface of a sufficiently massive star and its photograph is a further
problem of interest. A more substantial extension is the construction of the
photograph function in a rotating spacetime. The Kerr metric admits a fourth
conserved quantity, the Carter constant, and the orbits are no longer symmetric
about a perihelion, so that the relation between the two endpoints of a ray must
be treated differently; obtaining the analogous end-to-end function, and
determining whether the spin parameter can be inferred from the precise image
positions, would be of considerable interest.

% ===========================================================================
\appendix
% ===========================================================================

\section{Null geodesic equations in Schwarzschild spacetime}
\label{app:geodesics}

The Schwarzschild metric is well known and can be written as
\begin{equation}
\begin{aligned}
ds^{2}={}&-\left(1-\frac{2M}{r}\right)dt^{2}
     +\left(1-\frac{2M}{r}\right)^{-1}dr^{2}\\
     &+r^{2}\left(d\theta^{2}+\sin^{2}\theta\,d\phi^{2}\right),
\end{aligned}
\label{eq:metric}
\end{equation}
where $M$ is the mass, the event horizon is at $r_{h}=2M$, and the photon sphere
is at $r_{\mathrm{ph}}=3M$. Null geodesics satisfy the geodesic equation
\begin{equation}
\frac{d^{2}x^{\mu}}{d\lambda^{2}}
+\Gamma^{\mu}_{\rho\sigma}\frac{dx^{\rho}}{d\lambda}\frac{dx^{\sigma}}{d\lambda}=0 .
\label{eq:geo-eq}
\end{equation}
Because of time independence and spherical symmetry there are four conserved
quantities: the energy $E$, the total angular momentum $L$, the $z$ component of
the angular momentum $P_{\phi}$, and the mass $m=0$ of the photon, which are
\begin{align}
E &= \left(1-\frac{2M}{r}\right)\frac{dt}{d\lambda}, \label{eq:Edef} \\[2pt]
L^{2} &= r^{4}\left[\left(\frac{d\theta}{d\lambda}\right)^{2}
        +\sin^{2}\theta\left(\frac{d\phi}{d\lambda}\right)^{2}\right],
        \label{eq:Ldef} \\[2pt]
P_{\phi} &= r^{2}\sin^{2}\theta\,\frac{d\phi}{d\lambda}, \label{eq:Pdef} \\[2pt]
0 &= g_{\mu\nu}\frac{dx^{\mu}}{d\lambda}\frac{dx^{\nu}}{d\lambda}. \label{eq:nullcond}
\end{align}
With these quantities the geodesic equations~\eqref{eq:geo-eq} reduce to a
system of first-order ordinary differential equations,
\begin{align}
\frac{dt}{d\lambda} &= \frac{Er}{r-2M}, \label{eq:dt} \\[2pt]
\frac{dr}{d\lambda} &= \pm\sqrt{E^{2}-\frac{L^{2}}{r^{2}}
                       \left(1-\frac{2M}{r}\right)}, \label{eq:dr} \\[2pt]
\frac{d\theta}{d\lambda} &= \pm\frac{\sqrt{L^{2}-P_{\phi}^{2}\csc^{2}\theta}}{r^{2}},
                       \label{eq:dtheta} \\[2pt]
\frac{d\phi}{d\lambda} &= \frac{P_{\phi}}{r^{2}\sin^{2}\theta}, \label{eq:dphi}
\end{align}
where the $+$ and $-$ branches represent outgoing and incoming rays,
respectively.

We set up two coordinate systems, the celestial coordinate and the apparent
coordinate, as shown in Fig.~\ref{fig:coordinates}. The celestial coordinate is
defined as three-dimensional Cartesian coordinates $(x,y,z)$ with the black hole
at the origin $B$, and can be transformed into spherical coordinates
$(r,\theta,\phi)$ through
\begin{equation}
x=r\sin\theta\cos\phi,\qquad
y=r\sin\theta\sin\phi,\qquad
z=r\cos\theta .
\label{eq:cart}
\end{equation}
The apparent coordinate is a two-dimensional Cartesian coordinate $(R,Z)$ with
$x\parallel X$, which can be transformed into polar coordinates $(R,\Theta)$
through $X=R\sin\Theta$ and $Z=R\cos\Theta$.

The observer (or camera) is placed at $(x=0,\,y=-l\sin\Psi,\,z=l\cos\Psi)$ in the
celestial coordinate, that is, at $(r=l,\theta=\Psi,\phi=-\pi/2)$, or
equivalently at the origin $O$ of the apparent coordinate. The apparent
coordinates are defined on the tangent plane of a sphere of radius $l$. A point
located at $(R,\Theta)$ in the apparent coordinate has celestial coordinates
$(x,y,z)$ satisfying
\begin{equation}
\begin{aligned}
x&=X,\qquad y=-l\sin\Psi+Z\cos\Psi,\\
z&=l\cos\Psi+Z\sin\Psi .
\end{aligned}
\label{eq:apparent-cart}
\end{equation}
These relations can also be written in spherical coordinates as
\begin{equation}
\begin{aligned}
r&=\sqrt{l^{2}+R^{2}},\\
\cos\theta&=\frac{l\cos\Psi+R\cos\Theta\sin\Psi}{\sqrt{l^{2}+R^{2}}},\\
\tan\phi&=\frac{-l\sin\Psi+R\cos\Theta\cos\Psi}{R\sin\Theta}.
\end{aligned}
\label{eq:apparent-sph}
\end{equation}

Assuming that the observer is at a large distance from the source, $l\gg1$, the
light reaching the apparent coordinate has propagated over a sufficiently long
path and therefore forms a beam of parallel rays perpendicular to the apparent
coordinate plane. Thus a ray arriving at $(R_{a},\Theta_{a})$ with celestial
coordinates $(r_{a},\theta_{a},\phi_{a})$ perpendicular to the apparent plane
has initial four-velocity
\begin{equation}
\begin{aligned}
\frac{dt_{a}}{d\lambda}&=1,\qquad
\frac{dx_{a}}{d\lambda}=0,\\
\frac{dy_{a}}{d\lambda}&=\sin\Psi,\qquad
\frac{dz_{a}}{d\lambda}=-\cos\Psi .
\end{aligned}
\label{eq:four-vel}
\end{equation}
As for the time coordinate, it satisfies $t_{a}\propto\lambda$, so that the
affine parameter may simply be identified with the observer's time $t_{a}$. In
what follows we no longer consider time. Since the spherical coordinates are
related to the Cartesian ones by Eqs.~\eqref{eq:cart}, one can derive
\begin{align}
\frac{dx_{a}}{d\lambda} &= \frac{dr_{a}}{d\lambda}\sin\theta_{a}\cos\phi_{a}
  +\frac{d\theta_{a}}{d\lambda}r_{a}\cos\theta_{a}\cos\phi_{a}\notag\\
  &\quad -\frac{d\phi_{a}}{d\lambda}r_{a}\sin\theta_{a}\sin\phi_{a}, \label{eq:dca1}\\[2pt]
\frac{dy_{a}}{d\lambda} &= \frac{dr_{a}}{d\lambda}\sin\theta_{a}\sin\phi_{a}
  +\frac{d\theta_{a}}{d\lambda}r_{a}\cos\theta_{a}\sin\phi_{a}\notag\\
  &\quad +\frac{d\phi_{a}}{d\lambda}r_{a}\sin\theta_{a}\cos\phi_{a}, \label{eq:dca2}\\[2pt]
\frac{dz_{a}}{d\lambda} &= \frac{dr_{a}}{d\lambda}\cos\theta_{a}
  -\frac{d\theta_{a}}{d\lambda}r_{a}\sin\theta_{a}. \label{eq:dca3}
\end{align}
Combining Eqs.~\eqref{eq:four-vel} with
Eqs.~\eqref{eq:dca1}--\eqref{eq:dca3}, we obtain
\begin{align}
\frac{dr_{a}}{d\lambda} &= -\frac{l}{\sqrt{l^{2}+R_{a}^{2}}}, \label{eq:drinit}\\[4pt]
\frac{d\theta_{a}}{d\lambda} &=
  \frac{R_{a}\left(R_{a}\cos\Psi-l\cos\Theta_{a}\sin\Psi\right)}
       {(l^{2}+R_{a}^{2})\sqrt{\Delta_{\theta}}}, \label{eq:dthetainit}\\[4pt]
\frac{d\phi_{a}}{d\lambda} &=
  \frac{R_{a}\sin\Theta_{a}\sin\Psi}{\Delta_{\theta}},
  \label{eq:dphiinit}
\end{align}
where
\begin{equation}
\Delta_{\theta}\equiv
\left(R_{a}\cos\Theta_{a}\cos\Psi-l\sin\Psi\right)^{2}
+R_{a}^{2}\sin^{2}\Theta_{a}.
\label{eq:DeltaTheta}
\end{equation}
where Eqs.~\eqref{eq:apparent-sph} have been used to express
$(r_{a},\theta_{a},\phi_{a})$ in terms of $(R_{a},\Theta_{a})$ on the
right-hand side.

At the same time, the null geodesics must satisfy
Eqs.~\eqref{eq:dt}--\eqref{eq:dphi}. Substituting
$(r_{a},\theta_{a},\phi_{a})$ and considering the ingoing branch, one obtains
\begin{align}
\frac{dr_{a}}{d\lambda} &= -\sqrt{E^{2}-\frac{L^{2}\left(l^{2}+R_{a}^{2}\right)-2M}
                        {\left(l^{2}+R_{a}^{2}\right)^{3/2}}}, \label{eq:drgeo}\\[4pt]
\frac{d\theta_{a}}{d\lambda} &=
 -\frac{1}{l^{2}+R_{a}^{2}}
  \sqrt{L^{2}-\frac{P_{\phi}^{2}\left(l^{2}+R_{a}^{2}\right)}{\Delta_{\theta}}},
 \label{eq:dthetageo}\\[4pt]
\frac{d\phi_{a}}{d\lambda} &=
  \frac{P_{\phi}}{\Delta_{\theta}}. \label{eq:dphigeo}
\end{align}
The initial velocities given by Eqs.~\eqref{eq:drinit}--\eqref{eq:dphiinit} and
by Eqs.~\eqref{eq:drgeo}--\eqref{eq:dphigeo} are expected to coincide.
Comparing them gives
\begin{equation}
\begin{aligned}
E&=\sqrt{1-\frac{2MR_{a}^{2}}{\left(l^{2}+R_{a}^{2}\right)^{3/2}}},\qquad
L=R_{a},\\
P_{\phi}&=R_{a}\sin\Theta_{a}\sin\Psi .
\end{aligned}
\end{equation}
Thus the conserved quantities $E$, $L$ and $P_{\phi}$ are determined by the
apparent-coordinate position $(R_{a},\Theta_{a})$. Substituting them into
Eqs.~\eqref{eq:drgeo}--\eqref{eq:dphigeo} and writing $u=r^{-1}$, one recovers
Eqs.~\eqref{eq:eom-u}--\eqref{eq:eom-phi}, the subscript $a$ having been
dropped.

% ===========================================================================
\section{Exact solutions}
\label{app:solutions}
% ===========================================================================

In this appendix we solve the first-order ordinary differential
equations~\eqref{eq:eom-u}--\eqref{eq:eom-phi} exactly. For simplicity we
consider light rays that propagate freely, without requiring a light source. Not
every ray is able to reach the apparent plane. By the time-reversal symmetry of
light propagation it is convenient to suppose that a fictitious source exists at
each point of the apparent plane, emitting light towards the black hole. The
resulting trajectories fall into three categories, according to the roots of
$f(u)=0$:

\begin{enumerate}
\item Rays that plunge into the black hole. This corresponds to the case in
which $f(u)=0$ has one real root smaller than zero and two imaginary roots.

\item Rays that first reach the perihelion $r_{p}$ and are then scattered. This
corresponds to the case in which $f(u)=0$ has three distinct real roots
$e_{1}<0<e_{2}<e_{3}$, the perihelion being $r_{p}=1/e_{2}$.

\item Rays that neither reach the source nor fall into the hole, but instead
tend asymptotically to the photon sphere $r=3M$. This is the critical case,
satisfying $f(u)=0$ and $f'(u)=0$, for which $f(u)=0$ has two distinct real
roots with $e_{1}<0<e_{2}=e_{3}$, where $e_{2}=e_{3}=1/3M$.
\end{enumerate}

Equations~\eqref{eq:eom-u}--\eqref{eq:eom-phi} can be rewritten as
\begin{equation}
\begin{aligned}
\int\frac{du}{\sqrt{f(u)}}&=\pm\int
\frac{d\theta}{\sqrt{1-\csc^{2}\theta\sin^{2}\Theta\sin^{2}\Psi}}\\
&=\pm\int\frac{\sin^{2}\theta\,d\phi}{\sin\Theta\sin\Psi}.
\end{aligned}
\label{eq:separated}
\end{equation}
For case (2) the first of these integrals can be evaluated by means of the
Biermann--Weierstrass theorem.

\begin{theorem}[Biermann--Weierstrass]
Let
\begin{equation}
f(x)=a_{4}x^{4}+4a_{3}x^{3}+6a_{2}x^{2}+4a_{1}x+a_{0}
\end{equation}
be a quartic polynomial in $x$, and denote its invariants by
\begin{align}
g_{2} &= a_{0}a_{4}-4a_{1}a_{3}+3a_{2}^{2}, \label{eq:g2}\\
g_{3} &= a_{0}a_{2}a_{4}+2a_{1}a_{2}a_{3}-a_{3}^{2}-a_{1}^{2}a_{4}-a_{0}a_{3}^{2}.
\label{eq:g3}
\end{align}
Let
\begin{equation}
z(x)=\int_{x_{0}}^{x}\frac{dx'}{\sqrt{f(x')}},
\end{equation}
where $x_{0}$ is a constant satisfying $f(x_{0})\neq0$. Then
\begin{equation}
\begin{aligned}
x={}&\!x_{0}+\Big\{-\sqrt{f(x_{0})}\,\wp'(z)
 +\tfrac{1}{12}f'(x_{0})\big[\wp(z)-\tfrac{1}{24}f''(x_{0})\big] \\
 &+\tfrac{1}{24}f(x_{0})f'''(x_{0})\Big\}
 \Big/\Big\{2\big[\wp(z)-\tfrac{1}{24}f''(x_{0})\big]^{2} \\
 &-\tfrac{1}{48}f(x_{0})f^{(4)}(x_{0})\Big\},
\end{aligned}
\label{eq:bw-x}
\end{equation}
and
\begin{equation}
\begin{aligned}
\wp(z)={}&\frac{\sqrt{f(x)f(x_{0})}+f(x_{0})}{2(x-x_{0})^{2}}\\
      &+\frac{f'(x_{0})}{4(x-x_{0})}+\frac{f''(x_{0})}{24},
\end{aligned}
\label{eq:bw-p}
\end{equation}
\begin{equation}
\begin{aligned}
\wp'(z)={}&-\left[\frac{f(x)}{(x-x_{0})^{3}}
        -\frac{f'(x)}{4(x-x_{0})^{2}}\right]\sqrt{f(x_{0})} \\
 &-\left[\frac{f(x_{0})}{(x-x_{0})^{3}}
        -\frac{f'(x_{0})}{4(x-x_{0})^{2}}\right]\sqrt{f(x)},
\end{aligned}
\label{eq:bw-pprime}
\end{equation}
where $\wp(z;g_{2},g_{3})$ is the Weierstrass function with the invariants
\eqref{eq:g2} and \eqref{eq:g3}.
\end{theorem}

Here $f(u)=a_{4}u^{4}+4a_{3}u^{3}+6a_{2}u^{2}+4a_{1}u+a_{0}$ with
\begin{equation}
\begin{aligned}
a_{4}&=a_{1}=0,\qquad a_{3}=\frac{M}{2},\qquad a_{2}=-\frac{1}{6},\\
a_{0}&=\frac{1}{R^{2}}\left(1-\frac{2MR^{2}}{(l^{2}+R^{2})^{3/2}}\right).
\end{aligned}
\label{eq:coefficients}
\end{equation}
The solution for $u$ can then be written as
\begin{equation}
\begin{aligned}
 u(z)&=u_i\\
 &+\frac{-\sqrt{f(u_i)}\wp'(z)+(3M u_i^2-u_i)\left[ \wp(z)-\frac{6Mu_i-1}{12}\right]+\frac{M}{2}f(u_i)}{2\left[ \wp(z)-\frac{6Mu_i-1}{12}\right]^2}.
\end{aligned}
\label{eq:usolution}
\end{equation}
where the Weierstrass function $\wp$ for the lattice with invariants $g_{2}$,
$g_{3}$ satisfies
\begin{equation}
(\wp')^{2}=4\wp^{3}-g_{2}\wp-g_{3},
\label{eq:wpode}
\end{equation}
and
\begin{equation}
\begin{aligned}
g_{2}&=\frac{1}{12},\\
g_{3}&=\frac{1}{216}-\frac{M^{2}}{4R^{2}}
      \left(1-\frac{2MR^{2}}{(l^{2}+R^{2})^{3/2}}\right).
\end{aligned}
\label{eq:invariants-app}
\end{equation}

Direct integration of
\begin{equation}
z=\pm\int_{\theta_{i}}^{\theta}
\frac{d\theta}{\sqrt{1-\csc^{2}\theta\sin^{2}\Theta\sin^{2}\Psi}}
\label{eq:thetaintegral}
\end{equation}
gives the solution for $\theta$,
\begin{equation}
\begin{aligned}
z=\pm\Bigg[&\arcsin\!\frac{\sin\theta}{\sqrt{1-\sin^{2}\Theta\sin^{2}\Psi}}\\
       &-\arcsin\!\frac{\sin\theta_{i}}{\sqrt{1-\sin^{2}\Theta\sin^{2}\Psi}}\Bigg].
\end{aligned}
\label{eq:thetasolution}
\end{equation}
Similarly, one obtains the solution for $\phi$,
\begin{equation}
\begin{aligned}
\phi-\phi_{i}={}&\mp\left[
 \arctan\!\frac{\sin\Theta\sin\Psi\cos\theta}
             {\sqrt{1-\cos^{2}\theta-\sin^{2}\Theta\sin^{2}\Psi}}
\right.\\
&\left.-\arctan\!\frac{\sin\Theta\sin\Psi\cos\theta_{i}}
             {\sqrt{1-\cos^{2}\theta_{i}-\sin^{2}\Theta\sin^{2}\Psi}}\right],
\end{aligned}
\label{eq:phisolution}
\end{equation}
or equivalently
\begin{equation}
\cos\theta=\frac{\sqrt{1-\sin^{2}\Theta\sin^{2}\Psi}\,
                  \tan\left(A\mp(\phi-\phi_{i})\right)}
                {\sqrt{\sin^{2}\Theta\sin^{2}\Psi
                  +\tan^{2}\left(A\mp(\phi-\phi_{i})\right)}},
\label{eq:costheta}
\end{equation}
with
\begin{equation}
A=\arctan\!\frac{\sin\Theta\sin\Psi\cos\theta_{i}}
            {\sqrt{1-\cos^{2}\theta_{i}-\sin^{2}\Theta\sin^{2}\Psi}} .
\end{equation}

\begin{acknowledgments}
This work is partially supported by the National Natural Science Foundation of
China (Grant No. 12505071) and by the Research Foundation of the Education
Bureau of Hunan Province, China (Grant No. 25B0635). This work is also
partially supported by the Hunan Provincial Department of Education Key
Laboratory of Information Detection and Intelligent Processing Technology, and
by the Applied Characteristic Subject of Hunan Province "Electronic Science and
Technology".
\end{acknowledgments}

\end{document}